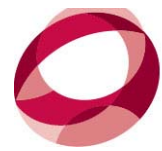

# Advanced Cyberinfrastructure for Science, Engineering, and Public Policy[1]

Vasant G. Honavar, Katherine Yelick, Klara Nahrstedt, Holly Rushmeier, Jennifer Rexford, Mark D. Hill, Elizabeth Bradley, and Elizabeth Mynatt
June 2017

**Abstract**
Progress in many domains increasingly benefits from our ability to view the systems through a computational lens, i.e., using computational abstractions of the domains; and our ability to acquire, share, integrate, and analyze disparate types of data. These advances would not be possible without the advanced data and computational cyberinfrastructure and tools for data capture, integration, analysis, modeling, and simulation. However, despite, and perhaps because of, advances in "big data" technologies for data acquisition, management and analytics, the other largely manual, and labor-intensive aspects of the decision making process, e.g., formulating questions, designing studies, organizing, curating, connecting, correlating and integrating cross-domain data, drawing inferences and interpreting results, have become the rate-limiting steps to progress. Advancing the capability and capacity for evidence-based improvements in science, engineering, and public policy requires support for (1) computational abstractions of the relevant domains coupled with computational methods and tools for their analysis, synthesis, simulation, visualization, sharing, and integration; (2) cognitive tools that leverage and extend the reach of human intellect, and partner with humans on all aspects of the activity; (3) nimble and trustworthy data cyber-infrastructures that connect, manage a variety of instruments, multiple interrelated data types and associated metadata, data representations, processes, protocols and workflows; and enforce applicable security and data access and use policies; and (4) organizational and social structures and processes for collaborative and coordinated activity across disciplinary and institutional boundaries.

**Motivating Research Challenges**
The emergence of "big data" offers unprecedented opportunities for not only accelerating progress, but it also dramatically transforms science, engineering and public policy. For example, advances in sequencing, imaging, and online text processing, coupled with domain-specific computational abstractions, and new methods and tools for data integration, analysis, and modeling, are enabling: biologists to gain insights into how living systems adapt and thrive;

---

[1] Adapted in part from the CCC response to NSF Request for Information (RFI) on Future Needs for Advanced Cyberinfrastructure to Support Science and Engineering Research (NSF CI 2030), and based in part on the contents of: Honavar V., Hill M., & Yelick K. (2016). *Accelerating Science: A Computing Research Agenda*: A white paper prepared for the Computing Community Consortium committee of the Computing Research Association. http://cra.org/ccc/wp-content/uploads/sites/2/2016/02/Accelerating-Science-Whitepaper-CCC-Final2.pdf

neuroscientists to uncover how brains adapt and learn; health scientists to personalize treatments and interventions to optimize health outcomes; economists to understand markets; education researchers to personalize curricula and pedagogy to optimize learning outcomes; social scientists to study why organizations, societies, and cultures succeed or fail; material scientists to develop new materials with properties not seen before (materials genomics initiative).

Progress in many areas of human endeavor is increasingly enabled by our ability to view natural and manmade systems through the computational lens, i.e., using computational abstractions of the domains; and our ability to acquire, share, integrate, and analyze disparate types of data. These advances would not be possible without the advanced data and computational infrastructure and tools for data capture, integration, analysis, modeling, and simulation. However, despite, and perhaps because of, advances in "big data" technologies for data acquisition, management and analytics, the other largely manual, and labor-intensive aspects the scientific process, e.g., formulating questions, designing, prioritizing and executing studies, generating hypotheses, organizing, curating, coordinating, correlating and integrating data, drawing inferences, interpreting results, and evaluating models, have become the rate-limiting steps in scientific progress, human knowledge and decision making in science, engineering, and public policy. Advancing the capability and capacity for evidence-based decision making in science, engineering, and public policy requires support for *computational abstractions* of relevant domains, coupled with computational methods and tools for their analysis, synthesis, simulation, visualization, sharing, and integration; *cognitive tools* that leverage and extend the reach of human intellect, and partner with humans on all aspects of activity; *agile and trustworthy data cyber-infrastructures* that connect, manage diverse instruments; *representations, processes, protocols, workflows* that embody computational abstractions of the activities in question; and *organizational and social structures and processes* for collaboration and coordination that transcends disciplinary and institutional boundaries.

Consider, for example, the task of identifying a question for investigation in a domain of inquiry, e.g., the Life Sciences. This is a non-trivial task that requires a good grasp of the current state of knowledge, the expertise and skills needed, the instruments of observation available, the experimental manipulations that are possible, the data analysis and interpretation tools available, etc. Understanding the current state of knowledge requires mastery of the relevant scientific literature, which much like many other kinds of "big data" is growing at an exponential rate. The sheer volume and the rate of growth of scientific literature makes it impossible for a scientist to keep up with advances that might have a bearing on the questions being pursued in one's laboratory. The magnitude of this challenge is further compounded by the fact that many scientific investigations increasingly need to draw on data from a multitude of databases (e.g., Genbank, Protein Data Bank, etc. in the life sciences) and expertise and results from multiple disciplines.

As a second example, consider the task of designing a new material with some desired properties. This requires a careful exploration of the space of possible material compositions and manufacturing processes that could yield the material with the desired properties, their relative cost, risk, and feasibility, in the context of all that is known in the relevant areas of science and engineering. However, the absence of shared data storage and management services and agreed upon shared vocabularies, metadata standards and ontologies presents a significant hurdle to data and metadata acquisition from scientific instruments, data curation and discovery, and data exchange with other cross-domain scientists. Absence of understanding of different upgrade time scales for scientific instruments, computing, communication, security hardware and software causes slowdown in data acquisition, hence their fast integration into the overall scientific process. Furthermore, there is an absence of efforts to combine data from simulations, instruments' experiments, and observations, hence to generate new materials with properties that we have not see before.

As a third example, consider the task of understanding the complex interactions among genetic, environmental, behavioral, socio-demographic, and other factors in determining the risks of different health outcomes (e.g., cancer, cardiovascular disease, diabetes), and ultimately, designing risk-mitigating interventions. Some of the data, essential for such analyses, e.g., electronic health records, is highly sensitive in nature. Consequently, any analyses of such data must conform to applicable data access and use policies while maximizing the scientific utility of the data and the analyses. Cyberinfrastructure and data governance processes that support such analyses are largely lacking.

As a fourth example, consider the problem of protecting critical national infrastructures, for information and communication, energy production and transmission, food supply chain, healthcare delivery, among others. By its very nature, this problem requires integration of data, information, and decisions across multiple, relatively autonomous, organizations and stakeholders that operate under different legal and policy regimes. For example, in the cybersecurity domain, detecting, preventing, and responding to coordinated attacks on the infrastructure requires coordinated sharing of data, information, and decisions among autonomous entities across multiple sectors. This presents many challenges that remain to be addressed in terms of the supporting infrastructure.

The preceding examples are intended to be illustrative, and by no means exhaustive.

**Cyberinfrastructure Needed to Address the Research Challenge(s)**

We see the need for advances in cyberinfrastructure in several broad thematic areas:

**Computational abstractions of scientific, engineering, and public policy domains, and formal methods and tools for their analysis, integration, visualization, simulation, sharing, and discovery**

A key prerequisite for realizing the full potential of data and computation to advance the capability and capacity for evidence-based improvements in science, engineering, and public policy requires viewing the respective domains through a *computational lens*, that is, in terms of computational abstractions. Of particular interest are system-level, mechanistic, computational models of physical, biological, cognitive, and social systems that enable the integration of different processes into coherent and rigorous representations that can be analyzed, simulated, integrated, shared, validated against experimental data, and used to guide experimental investigations or interventions. These models must not only provide cross-levels of abstraction, but also disciplinary boundaries, to allow studies of complex interactions, e.g., those that couple food, energy, water, environment, and people.

**Hardware infrastructure**

High-end computational capabilities for modeling, simulation, data analysis and inference are very much needed. The need for such systems to support physical modeling and simulation problems continues to grow, due to the increasing complexity of scientific inquiry, addressing multiphysics problems and increasing dynamic range in space and time. This will drive the need for access to computational systems of the highest scales in order to be scientifically competitive. At the same time, there is an additional demand for high throughput computations, e.g., millions of materials modeling simulations to screen for materials with particular properties needed for batteries, electronics, and other applications. These simulations require systems similar to those at the largest scales, including tightly coupled parallel architectures with high performance floating point appropriately balanced with memory bandwidth, high speed interconnects (although at a smaller scale), and software to co-schedule parallel jobs across nodes.

High end and high throughput computations will need to be complemented by vast numbers of medium- and small-scale computing tasks suitable for initial exploration, work with more modest data sizes, etc. In many cases, these equally important tasks may be suitable for execution and storage made available by commercial Infrastructure-as-a-Service (IaaS) providers.

The growing size and complexity of data from simulations, experimental scientific instruments, and embedded sensors will also drive the need for computational infrastructure. With deep learning approaches alone, access to large amounts of computing to train these networks is as important as access to large data sets. Architectural choices for machine learning, graph analytics or other data analysis problems may be different than simulation, e.g., leveraging lower precision floating point or requiring low latency networks for walking over graphs, although today similar processor architectures (including GPUs, FPGA accelerators and heterogeneous processors architectures) are being used. An important feature of the use of computational lens in science, engineering, and public policy is that it involves integration of computation, data, models and prediction, suggesting that systems will need to handle diverse workloads to efficiently support the needed scientific activities and processes.

Data intensive problems will require advanced networking with unique demands for tools and systems to serve high-speed flows from large experiments, as well as on-demand computing and in-stream processing that are not suited to queue-based scheduling. Complex workflows will also benefit from containerized software and micro-service platforms, and storage models will need to evolve to address provenance, accessibility and sustained availability.

**Retrofitting tools for Instruments**

Scientific instruments, and other elements of physical infrastructure (e.g., in clean rooms) are often acquired to last for at least 10 years, with many instruments lasting for 15-20 years. The tools used for digital data acquisition are usually upgraded every 2-4 years depending on the manufacturer. Overall, the manufacturers are required to provide instrument's upgrades for 8 years. On the other hand, computing, communication and security hardware and software components (e.g., operating systems) are being upgraded/patched every 6 months on average. Hence, the next generation cyberinfrastructure for science needs to include considerations for retrofitting tools and technologies such as edge computing solutions to allow the instruments and their experimental scientific data to be connected in timely and trustworthy manner to advanced cloud and/or high performance computing infrastructures for further analysis, curation, and integration with simulation and modeling tools.

**Cognitive tools for scientists, engineers and decision makers**

The next generation cyberinfrastructure needs to provide a broad range of cognitive tools for scientists, engineers, and decision makers, i.e., computational tools that leverage and extend human intellect, and partner with humans on a broader range of tasks that make up a typical workflow (formulating a question, designing, prioritizing and executing experiments designed to answer the question, drawing inferences and evaluating the results, and formulating new questions, in a closed-loop fashion). That is, the cyberinfrastructure needs to support computational abstractions of various aspects of the activity; development of the computational artifacts (representations, processes, software) that embody such abstractions; and the integration of the resulting artifacts into collaborative human-machine systems to advance science, engineering and public policy, (by augmenting, and whenever feasible, replacing individual or collective human efforts). The resulting systems would need to close the loop from designing experiments to acquiring, curating, and analyzing data to generating and refining hypotheses back to designing new experiments or performing evidence-based interventions.

Particularly needed are cognitive tools for:
- Mapping the current state of knowledge in a discipline and identifying the major gaps;
- Generating and prioritizing questions that are ripe for investigation based on the current scientific priorities and the gaps in the current state of knowledge;
- Machine reading, including methods for extracting and organizing descriptions of experimental protocols, scientific claims, supporting assumptions, and validating scientific claims from scientific literature, and increasingly scientific databases and knowledge bases;
- Literature-based discovery, including methods for drawing inferences and generating hypotheses from existing knowledge in the literature (augmented with discipline-specific databases and knowledge bases of varying quality when appropriate), and ranking the resulting hypotheses;
- Expressing, reasoning with, and updating scientific arguments (along with supporting assumptions, facts, observations), including languages and inference techniques for managing multiple, often conflicting arguments, assessing the plausibility of arguments, their uncertainty and provenance;
- Observing and experimenting, including languages and formalisms for describing and harmonizing the measurement process and data models, capturing and managing data provenance, describing, quantifying the utility, cost, and feasibility of experiments, comparing alternative experiments, and choosing optimal experiments (in a given context);
- Navigating the spaces of hypotheses, conjectures, theories, and the supporting observations and experiments;
- Analyzing and interpreting the results of observations and experiments, including machine learning methods that: explicitly model the measurement process, including its bias, noise, resolution; incorporate constraints e.g., those derived from physics, into data-driven inference; generate hybrid approaches where physics-based models are then calibrated with data; close the gap between model builders and model users by producing models that are expressible in representations familiar to the disciplinary scientists;
- Synthesizing, in a principled manner, the findings, e.g., causal relationships from disparate experimental and observational studies (e.g., implications to human health of experiments with mouse models), and creating synthetic data

Because major activities in science, engineering, and public policy increasingly rely on collaboration across disciplinary as well as organizational boundaries, there is a need for data and computational infrastructure to support:

- Organizing and participating in team projects, including tools for decomposing tasks, assigning tasks, integrating results, incentivizing participants, and engaging large numbers of participants with varying levels of expertise and ability in the process;
- Collaborating, communicating, and forming teams with partners with complementary knowledge, skills, expertise, and perspectives on problems of common interest (including problems that span disciplinary boundaries or levels of abstraction, and call for collaboration across government, industry and academia);
- Creating and sharing of human understandable and computable representations of the relevant artifacts, including data, experiments, hypotheses, conjectures, models, theories, workflows, etc. across organizational and disciplinary boundaries;
- Documenting, sharing, reviewing, replicating, and communicating entire studies in the form of reproducible and extensible workflows (with provision for capturing data provenance);
- Automating the discovery, adaptation, and when needed, assembly of complex analytic workflows from available components;
- Communicating results of studies or investigations and integrating the results into the larger body of knowledge within or across disciplines or communities of practice;
- Tracking scientific progress, the evolution of scientific disciplines, and impact on science, engineering, or public policy.

**Trustworthy Data Cyberinfrastructure**

In many domains, e.g., healthcare, cyber-security, etc., increasingly relies on data that are subject to restrictions on access and use. Because important decisions rely on the results of data analytics, there is a growing realization for the need for transparency, accountability, and trustworthiness of the cyberinfrastructure and the analytic processes. Hence, we need:
- Computable data access and usage agreements that can be enforced within a secure cyberinfrastructure;
- Audit mechanisms that can be used to verify compliance with the applicable data access and use policies;
- Repositories of data and their usage agreements that can be adapted and reused in a variety of settings.
- Agile and secure computing and network services and protocols that can adjust and protect different types and ages of instruments (i.e., longevity of cloud). For example, many scientific instruments (e.g., microscope), or components of physical infrastructure (e.g., power plants) that produce data are acquired to last a decade or longer, and hence the software used to operate them are often purchased and upgraded on a schedule that can be quite different from that used for updating the computing and networking software (operating system, security, network services) across the enterprise.

- Access privileges that are responsive to the changing needs and roles of individuals. For example, in the beginning of the scientific process, scientists are very protective of their data. However, after publishing the results of their study the same scientists may be open to publish and share data. In other settings, e.g., cybersecurity, the access privileges may need to change to accommodate the changing roles of, and relations among, the organizations or individuals in question.
- Distributed data management systems or virtual collaboratories that enable seamless sharing of data, computational resources, analysis tools, and results across disciplinary and organizational boundaries while ensuring compliance with applicable security as well as data access and use policies.
- Sustainable model for data and long-term preservation of both data and the software needed to make use of it.  Scientific data is often acquired at great cost. Data sets must be preserved to test alternative future interpretations, and to be extended and enriched by future data collection. Policies and mechanisms are needed for allowing stakeholders in collaborations to make appropriate copies of data critical to their applications for preservation. Policies and techniques are needed to preserve working copies of software and systems needed to read and process data, including emulators of software and data translators.
- Integration of computational and storage cyber-infrastructures that enable driving of computational simulations from acquired instrument data, and not just from models. Such integration would allow not only data-driven simulations and decision-making, but also learning and inference of realistic model functions from very large data sets that could be then used in computational simulators.
- Data provenance and other mechanisms for ensuring transparency of data analytics and decision making processes, detecting, and whenever possible, correcting for, implicit or explicit biases in the data as well as the algorithms used, that if left uncorrected, could adversely impact the integrity of, and public trust in, scientific, engineering, or public policy results obtained using advanced cyberinfrastructure.

Addressing real-world scientific, engineering, and public policy challenges requires a deep integration of knowledge, techniques, and expertise that transcend disciplinary as well as organizational boundaries, cyberinfrastructure in support of $21^{st}$ century science, engineering and public policy. Hence, the advanced cyberinfrastructure needs to support multi-disciplinary, interdisciplinary, and transdisciplinary teams that bring together:
- Scientists, engineers, and decision makers to develop computational abstractions of abstractions to support theoretical and experimental investigations, design and engineering activities, and decisions.
- Organizational and social scientists and cognitive scientists to study such teams at work, learn how best to organize and incentivize such teams to optimize their effectiveness;

- Scientists and engineers in one or more disciplines, computer and information scientists and engineers, organizational and social scientists, cognitive scientists, and others to design, implement, and study end-to-end systems that flexibly integrate the relevant cognitive tools into complex workflows to solve broad classes of problems in specific domains, e.g., improving population health, securing critical infrastructure, etc.

Training the 21$^{st}$ century scientists, engineers, and decision-makers who can both leverage and contribute to advances in cyberinfrastructure requires interdisciplinary graduate and undergraduate curricula and research based training programs to prepare:
- A diverse cadre of computer and information scientists and engineers with adequate knowledge of one or more scientific, engineering or public policy areas to design, construct, analyze and apply computational abstractions, cognitive tools, and end-to-end workflows in those disciplines;
- A new generation of natural, social, and cognitive science researchers, practitioners, and decision makers fluent in the use of computational abstractions and cognitive tools and advanced cyberinfrastructure to dramatically advance science, engineering and public policy.

Ensuring that the cyberinfrastructure investments lead to advances in the state-of-the-art in computational and data infrastructure for science, engineering, and public policy requires support for:
- Operational infrastructure based on the best available computing and information technology;
- Integrated data cyber-infrastructure that allows the sharing of data and metadata from simulations and experiments on scientific instruments;
- Experimental infrastructure to explore novel data, computing, and collaborative technologies and platforms, including the basic computer science and engineering advances needed to meet the needs of 21st century science;
- Datasets and tools for training and education that include data from successful as well as failed studies, including experiments and computational analyses.

*This material is based upon work supported by the National Science Foundation under Grant No. 1136993. Any opinions, findings, and conclusions or recommendations expressed in this material are those of the authors and do not necessarily reflect the views of the National Science Foundation.*

*For citation use: Honavar V., Yelick K., Nahrstedt K., Rushmeier H., Rexford J., Hill M., Bradley E., & Mynatt E. (2017). Advanced Cyberinfrastructure for Science, Engineering, and Public Policy.* http://cra.org/ccc/resources/ccc-led-whitepapers/